\documentclass[natbib=true,manuscript, screen]{acmart}
\AtBeginDocument{%
  }

\setcopyright{none}
\copyrightyear{2025-02}
\acmYear{2025}

\usepackage{xcolor}
\usepackage{hyperref}
\usepackage{listings}
\colorlet{punct}{red!60!black}
\definecolor{background}{HTML}{EEEEEE}
\definecolor{delim}{RGB}{20,105,176}
\colorlet{numb}{magenta!60!black}
\lstdefinelanguage{json}{
    basicstyle=\small\ttfamily,
    numbers=left,
    numberstyle=\scriptsize,
    stepnumber=1,
    numbersep=8pt,
    showstringspaces=false,
    breaklines=true,
    frame=lines,
    backgroundcolor=\color{background},
    literate=
     *{0}{{{\color{numb}0}}}{1}
      {1}{{{\color{numb}1}}}{1}
      {2}{{{\color{numb}2}}}{1}
      {3}{{{\color{numb}3}}}{1}
      {4}{{{\color{numb}4}}}{1}
      {5}{{{\color{numb}5}}}{1}
      {6}{{{\color{numb}6}}}{1}
      {7}{{{\color{numb}7}}}{1}
      {8}{{{\color{numb}8}}}{1}
      {9}{{{\color{numb}9}}}{1}
      {:}{{{\color{punct}{:}}}}{1}
      {,}{{{\color{punct}{,}}}}{1}
      {\{}{{{\color{delim}{\{}}}}{1}
      {\}}{{{\color{delim}{\}}}}}{1}
      {[}{{{\color{delim}{[}}}}{1}
      {]}{{{\color{delim}{]}}}}{1},
}

\begin{document}

\title[Evaluation of Sakana's AI Scientist]{Evaluating Sakana’s AI Scientist for Autonomous Research: Wishful Thinking or an Emerging Reality Towards ‘Artificial Research Intelligence’ (ARI)?}

\author{Joeran Beel}
\email{joeran.beel@uni-siegen.de}
\orcid{0000-0002-4537-5573}
\affiliation{%
  \institution{University of Siegen, \href{https://isg.beel.org}{Intelligent Systems Group} \& \href{https://Recommender-Systems.com}{Recommender-Systems.com} }
  \city{Siegen}
   \country{Germany}
}

\author{Min-Yen Kan}
\email{kanmy@comp.nus.edu.sg}
\orcid{0000-0001-8507-3716}

\affiliation{%
  \institution{National University of Singapore -- \href{https://wing.comp.nus.edu.sg/}{Web, Information Retrieval / Natural Language Processing Group (WING)}}
  \country{Singapore}
}

\author{Moritz Baumgart}
\email{moritz.baumgart@student.uni-siegen.de}
\orcid{0009-0007-1322-1450}
\affiliation{%
  \institution{University of Siegen}
  \city{Siegen}
   \country{Germany}
}

\renewcommand{\shortauthors}{Beel \& Kan et al.}

\begin{abstract}
\textbf{Abstract.} Recently, Sakana.ai introduced the AI Scientist, a system claiming to automate the entire research lifecycle and conduct research autonomously, a concept we term \textit{Artificial Research Intelligence} (ARI). Achieving ARI would be a major milestone toward Artificial General Intelligence (AGI) and a prerequisite to achieving Super Intelligence. The AI Scientist received much attention in the academic and broader AI community. A thorough evaluation of the AI Scientist, however, had not yet been conducted.

We evaluated the AI Scientist and found several critical shortcomings. The system's literature review process is inadequate, relying on simplistic keyword searches rather than profound synthesis, which leads to poor novelty assessments. In our experiments, several generated research ideas were incorrectly classified as novel, including well-established concepts such as micro-batching for stochastic gradient descent (SGD). The AI Scientist also lacks robustness in experiment execution—five out of twelve proposed experiments (42\%) failed due to coding errors, and those that did run often produced logically flawed or misleading results. In one case, an experiment designed to optimize energy efficiency reported improvements in accuracy while consuming more computational resources, contradicting its stated goal. Furthermore, the system modifies experimental code minimally, with each iteration adding only 8\% more characters on average, suggesting limited adaptability. The generated manuscripts were poorly substantiated, with a median of just five citations per paper—most of which were outdated (only five out of 34 citations were from 2020 or later). Structural errors were frequent, including missing figures, repeated sections, and placeholder text such as “Conclusions Here”. Hallucinated numerical results were contained in several manuscripts, undermining the reliability of its outputs.

Despite its limitations, the AI Scientist represents a significant leap forward in research automation. It produces complete research manuscripts with minimal human intervention, challenging conventional expectations of AI-generated scientific work. Many reviewers or university instructors conducting only a superficial assessment may struggle to distinguish its output from that of human researchers, demonstrating how far AI has progressed in mimicking academic writing and structuring scientific arguments. While the quality of its manuscripts currently aligns with that of an unmotivated undergraduate student rushing to meet a deadline, this level of autonomy in research generation is remarkable. More strikingly, it achieves this at an unprecedented speed and cost efficiency—our analysis indicates that generating a full research paper costs only \$6–\$15, with just 3.5 hours of human involvement. This is significantly faster than traditional human researchers. Given that AI research automation was nearly nonexistent just a few years ago, the AI Scientist marks a substantial milestone toward Artificial Research Intelligence (ARI), signalling the acceleration of AI-driven scientific discovery.

The AI Scientist also illustrates the urgent need for a discussion within the Information Retrieval (IR) and broader scientific communities. Whether and when ARI becomes a reality depends on how the academic and AI communities shape its development and governance. We propose concrete steps, including pilot projects and competitions, and standardized attribution frameworks such as research logs and markup languages. 

\end{abstract}




\maketitle

\section{Introduction}
In autumn 2024, Sakana.ai, a Tokyo-based start-up that has raised \$200 million in funding\footnote{\url{https://sakana.ai/series-a/}}, announced the ``AI~Scientist''. With the AI~Scientist, Sakana boldly promised the "beginning of a new era in scientific discovery" \cite{Sakana2024}. The open-source\footnote{\url{https://github.com/SakanaAI/AI-Scientist}} AI~Scientist is supposed to ``automate the entire research lifecycle''; i.e. it generates research ideas, designs and conducts experiments, analyzes results, writes research papers, and finally even reviews them — essentially automating the daily work of millions of researchers worldwide. According to Sakana, the AI~Scientist produces research papers for approximately \$15 each. Sakana acknowledges ``occasional flaws'' and explains further limitations in a pre-print manuscript \cite{lu2024aiscientistfullyautomated}. Yet, based on Sakana's released information, most readers will understand that the AI~Scientist appears to be a fully-functional system. Especially considering Sakana's claims that ``It is worth noting that the [AI~Scientist] can autonomously run the entire life cycle of machine learning research without any human intervention except for initial preparation''~\cite{Sakana.AI2024} and that the peer review system works ``with near-human accuracy'' \cite{Sakana2024}. 

Sakana’s AI~Scientist is not the only AI tool that aims to support or even replace significant aspects of scientific work. A rapidly expanding body of research explores AI’s role in scientific discovery, with recent preprints examining its impact on literature retrieval, idea generation, and automated experimental design. Several studies indicate that large language models (LLMs) can already generate research ideas comparable to those of human scientists \cite{Tyser2024,Ye2024,Su2024}, with some findings suggesting AI may even surpass human creativity \cite{Si2024,Radensky2025}. Google has framed AI as ushering in a “new era of scientific discovery” \cite{Manyika2024}, while specialized workshops such as ``Towards Agentic AI for Science'' at ICLR 2025 highlight the field’s growing momentum\footnote{\url{https://blog.iclr.cc/2025/02/03/workshops-at-iclr-2025}}. 

Notably, just one day before the submission of our current manuscript for review to a major IR conference, and hours before uploading it to arXiv, Google announced the \textit{AI Co-Scientist}\footnote{\url{https://blog.google/feed/google-research-ai-co-scientist/}}, indicating the emerging interest in autonomous research agents, or, as we term it ‘Artificial Research Intelligence’ (ARI), i.e. artificial intelligence that has not yet reached the level of general intelligence, but intelligence that is capable of conducting research that is indistinguishable from human research. In this context it is also interesting to note that \cite{Li2024} submitted a paper on AI-driven idea generation to ICLR 2025, which was rejected after an intense review process involving 45 discussion posts — described by the conference chair as ``filled with drama''\footnote{\url{https://openreview.net/forum?id=GHJzxPgFa6\&noteId=tN3UypVtcw}}. This case exemplifies the unease among researchers and reviewers who face the prospect of AI encroaching on their roles, raising questions about the future of human-led scientific inquiry. 
 
Despite the large body of work, the AI~Scientist is probably the tool that has made the boldest promises, yet. And, possibly because of this, it has gained the most attention, both within the academic community and among the broader AI enthusiast audience. As of early February 2025, Sakana's pre-print manuscript gained more than 100 citations\footnote{\url{https://scholar.google.com/scholar?cites=3682611450429091902}}, their GitHub repository received 8.9k stars and it was forked over 1.3k times. Prominent AI content creators on YouTube and X, each with hundreds of thousands of followers \cite{Berman2024,TheAIGRID2024}, enthusiastically discuss the AI~Scientist. 

Notably, the most positive reviewers of the AI~Scientist, e.g., \cite{Berman2024}, have not tested the system themselves but rely solely on Sakana’s information. Meanwhile, users in online forums and the GitHub issue trackers report difficulties in setting up the AI~Scientist\footnote{\url{https://www.reddit.com/r/learnmachinelearning/comments/1fuw2yb/has_anybody_gotten_sakana_ai_scientist_to_work/}} or encountering technical issues\footnote{\url{https://github.com/SakanaAI/AI-Scientist/issues}}. Other discussions\footnote{\url{https://spectrum.ieee.org/ai-for-science-2}} use the AI~Scientist’s release as a broader reflection on AI’s role in scientific discovery. Some reviews provide in-depth analyses\footnote{\url{https://medium.com/@nimritakoul01/evaluating-the-ai-scientist-63e419e575b8}}, examining its source code and generated papers. However, to the best of our knowledge, a comprehensive, independent assessment based on direct experimentation has yet to be conducted.

The AI~Scientist is a machine learning driven system, making it highly relevant to the machine learning research community. However, its impact extends beyond machine learning, with profound implications for the Information Retrieval (IR) community, which motivates us to write this 
 paper. While a recent SIGIR perspective paper \cite{Zhaisigir2024perspectice} provided valuable insights into how LLMs are transforming traditional IR tasks, we must also consider a broader question: What happens when LLMs move beyond assisting IR research and begin to conduct it autonomously? This possibility is no longer theoretical — AI systems are increasingly capable of generating research ideas, designing experiments, and even evaluating scientific work.

In this paper, we take the next step in this discussion by providing an independent, systematic assessment of the AI~Scientist’s capabilities, limitations, and trajectory. Our goal is not only to evaluate its functionality but to highlight its potential role in reshaping scientific inquiry, particularly within information retrieval. By analyzing its performance across idea generation, experimentation, and manuscript production and review, we offer a comprehensive perspective on what AI-driven research tools can — and cannot — achieve today and may achieve in the future. 

Our study focuses exclusively on Sakana’s AI~Scientist rather than other comparable tools due to its particularly bold claims and the significant attention it has garnered within the research community. While numerous AI-driven tools support various aspects of scientific research—including AI-assisted literature retrieval and summarization (Semantic Scholar, Elicit, Scite.ai), automated experiment design (IBM’s Watson Discovery), and research paper generation — the AI~Scientist stands out for its promise to fully automate the entire research lifecycle, from idea generation to peer review. This ambitious goal, combined with its open-source nature and extensive media coverage, makes it a uniquely compelling subject for independent evaluation. Other tools, while valuable in their respective domains, typically focus on specific tasks rather than claiming end-to-end automation of scientific inquiry. By concentrating on the AI~Scientist, we aim to provide a detailed and critical analysis of its capabilities and limitations without diluting our focus across multiple tools with varying objectives and architectures.

Our findings reveal a system that, while far from replacing human researchers, demonstrates the potential to automate significant parts of the research process. The AI~Scientist does not yet fulfill its promises, struggling with methodological soundness, experimental execution, and literature retrieval. However, these limitations are technical hurdles rather than fundamental barriers --- challenges that will likely be addressed as AI-driven research systems evolve.

For the IR community, the emergence of such tools presents both an opportunity and a challenge. On one hand, they offer new ways to automate literature retrieval, citation analysis, and experimental design, potentially advancing the field. On the other, they raise critical questions about the role of human researchers, the nature of scientific discovery, and the reliability and generalizability of AI-generated knowledge.

The time to engage with these technologies is now. Whether by integrating them into research workflows, critically evaluating their outputs, or contributing to their development, IR researchers must take an active role in shaping the future of AI-driven science. These systems are poised to transform how knowledge is produced and disseminated, and the IR community is uniquely positioned to lead this transformation.

\section{AI~Scientist: Functionality and Evaluation}
This section includes an overview of the features and architecture of the AI~Scientist, case study snippets, critiques and recommendations for use. 

\subsection{Setup and Installation}

All setup, coding and experiment execution were carried out by the third author of this manuscript, a third-year computer science Bachelor's student with basic machine learning knowledge but strong Python proficiency. A typical user of the AI~Scientist is likely to have greater expertise, such as that of a trained researcher or machine learning engineer. Thus, our setup time and usability assessments should be considered lower bounds.
We have made our experimental codebase that replicates our experimentation in this paper available on GitHub: \url{https://code.isg.beel.org/ISG-Siegen/the-ai-scientist-reproduced}. The repository includes a Singularity container definition file to facilitate full reproducibility and modifications in our experiments.

We installed the AI~Scientist on two computers and found the setup to be relatively straightforward, contrary to some reports on the Web \footnote{\url{https://www.reddit.com/r/learnmachinelearning/comments/1fuw2yb/has_anybody_gotten_sakana_ai_scientist_to_work/}}. The entire process took approximately five hours, including minor troubleshooting. For more details, we refer the interested reader to our GitHub repository.

We first tested and prototyped the AI~Scientist on a consumer laptop\footnote{HUAWEI MateBook D 16, Windows 11, Intel Core i5-12450H with integrated graphics, 16GB RAM (3733 MT/s)}. The main experiments were then conducted on our university's computing cluster\footnote{Single node with 2 AMD EPYC 7452 CPUs (32 cores, 2.35-3.35 GHz, 128 MB cache) and 256 GB DDR4 RAM (3200 MHz), running Rocky Linux 8.8}. Although the AI~Scientist provides an experimental pipeline optimized for deep learning with PyTorch and an NVIDIA GPU (CUDA), our experiments used a simple recommendation algorithm, FunkSVD, that runs on a CPU. 

The AI~Scientist requires a foundation large language model (LLM) to function. At the time of our experiments, it supported OpenAI's \texttt{gpt-4o-2024-05-13}, multiple models from Anthropic's \texttt{Claude} series, DeepSeek Coder V2, and Meta's \texttt{Llama 3.1}. We opted for \texttt{gpt-4o-2024-05-13}. Since then, Sakana has expanded support to additional models.


\subsection{Preparing for the experiments}
We initially assumed the AI~Scientist could autonomously conduct research based solely on a prompt. However, it requires a user-defined ``template,'' which significantly limits the autonomy of the AI Scientist. Such a template consist of several elements, described in the following.
\paragraph{1. Goal and Research Direction}
The research goal is specified via a prompt file (\textit{prompts.json}). For our experiments, we focused on Green Recommender Systems \cite{Beel2024e}, a research topic which aims to reduce the carbon footprint of recommender systems and its associated machine learning algorithms. We gave the prompt as: \footnote{All prompts are shortened and rephrased for brevity; the originals are available in our GitHub repository.}:

\begin{lstlisting}[caption={The general task provided as a JSON prompt},label=lst:task_json_code,language=JSON]   
{
  "task_description": "The attached code trains and evaluates the FunkSVD algorithm with stochastic gradient descent for recommender systems. Your goal is to find novel ways to optimize energy-efficiency. "
}
\end{lstlisting}

\paragraph{2. Experimental Pipeline}

The AI Scientist requires a Python-based experimental pipeline, consisting of an \texttt{experiment.py}, which defines dataset usage and model training as well as a \texttt{plot.py}, which handles result visualization. We implemented FunkSVD for collaborative filtering, training and evaluating it on an 80/10/10 split on MovieLens-100k (RMSE metric), with results visualized via line charts. This minimalist setup runs on a CPU, but larger datasets and complex models would likely increase compute costs.  It is important to note that the AI Scientist requires this pipeline in a special format. Users cannot simply input any Python code but would have to adjust their existing code to be able to be used by the AI Scientist.

\paragraph{3. Seed Ideas}
The AI~Scientist further requires user-provided seed ideas. We supplied two:  

\begin{enumerate}
    \item \textbf{Adaptive Learning Rates for SGD.} Aims to accelerate convergence, reducing computational cost and energy use. Though not novel, this ideas demonstrates our intended research direction.  
    \item \textbf{E-Fold Cross-Validation.} An alternative to k-fold cross-validation that dynamically selects an optimal \( e \) to balance computational efficiency and reliability. This idea is also not entirely novel, but was proposed only recently\cite{Mahlich2024,Beel2024c,Baumgart2024}. 
\end{enumerate}

\noindent The seed ideas must include values for ``interestingness,'' ``feasibility,'' and ``novelty.'' However, we observed that these values had no apparent impact on the AI~Scientist’s processing.

\begin{lstlisting}[caption={e-fold Cross Validation; one of the two seed ideas},label=lst:seed_ideas_json_code,language=JSON]   
{
    "Name": "e_fold_cv",
    "Title": "E-Fold Cross-Validation for Model Performance Estimation",
    "Experiment": "Develop an alternative to k-fold cross-validation named e-fold cross-validation. Instead of a static k it uses an intelligently chosen or dynamically adjusted paramter e to optimize the number of folds and to minimize computational energy while maintaining reliable model performance estimates.",
    "Interestingness": 8,
    "Feasibility": 7,
    "Novelty": 7
  }

\end{lstlisting}

\paragraph{4. \LaTeX Templates}
The AI~Scientist generates research papers using predefined LaTeX templates. We used the default templates provided by Sakana without modification.

\subsection{Idea Generation}
Once provided with required input, the AI~Scientist generated the following ten research ideas:

\begin{tabular}{rp{7.5cm}}  
    1. & Factor pruning for energy-efficient matrix factorization \\  
    2. & Dynamic factor adjustment in matrix factorization \\  
    3. & Green metrics for sustainable recommender systems \\  
    4. & Quantization techniques for energy-efficient SGD \\  
    5. & Time-aware early stopping  \\  
    6. & Interaction-priority SGD for energy efficiency \\  
    7. & Micro-batching for energy-efficient SGD \\  
    8. & Sparse-aware SGD for matrix factorization \\  
    9. & Cluster-aware SGD for matrix factorization \\  
    10. & Hybrid matrix factorization 
\end{tabular}

\noindent Each generated idea includes a brief description and three AI~Scientist judged numerical scores (1--10) for Interestingness, Feasibility, and Novelty, along with a binary ``novel'' label (True/False). However, while the \textit{novel} label affects its workflow, the other scores appear arbitrarily assigned and are not used in further processing. To determine novelty, the AI~Scientist queries the Semantic Scholar API, retrieving up to 10 search results per query. The system extracts search result metadata (title, authors, venue, year, citations, and abstracts) and can iterate up to 10 times before deciding. If no clear matches are found, it assigns \texttt{novel=True}; otherwise, it may classify the idea as non-novel. 

Despite mimicking a literature review, this approach is unreliable. In our experiments, the AI~Scientist classified all 10 generated ideas and both seed ideas as novel\footnote{In prior tests, the AI~Scientist occasionally flagged some ideas as non-novel, showing that it does not always default to a ``novel'' classification.}, despite some being well-documented. For instance, micro-batching for SGD (Idea~7) is a known technique \cite{jain2018parallelizing}, as are adaptive learning rates (seed idea), hybrid matrix factorization (idea 10) and e-fold cross validation (seed idea). This misclassification likely results from the AI~Scientist's reliance on keyword matching rather than any deeper synthesis of research. While some ideas may be novel within recommender systems, many simply apply existing techniques (e.g., pruning) to matrix factorization, offering incremental rather than groundbreaking improvements.

\subsection{Conducting Experiments}

The AI~Scientist executes research ideas by modifying the user-provided experimental pipeline using \texttt{Aider}\footnote{\url{https://Aider.chat/}}, an LLM-driven coding assistant that iterates on \texttt{experiment.py} up to five times. This reliance on \texttt{Aider} adds complexity, as errors in one system propagate to the other. While the AI~Scientist does introduce variation, it remains highly dependent on the provided framework, reworking the predefined templates rather than devising novel methodologies. 

Our code analysis suggests minimal changes per iteration. Our template had 6,260 characters of code (255 lines), with \texttt{Aider} adding, on average, 529 characters (+8\%) in the first iteration, followed by 118, 83, 66, and 21 characters in subsequent iterations (see Table~\ref{tab:character-changes}).

\begin{table*}
    \centering
    \begin{tabular}{|c|c|c|c|c|c|l|} \hline  
         &  &  \multicolumn{5}{|c|}{\textbf{Coding Iteration}}\\ \hline  
         \textbf{ID}
&  \textbf{Idea}
&  \textbf{1}
&  \textbf{2}
&  \textbf{3}
&  \textbf{4}
&\textbf{5}
\\ \hline  
         Seed 1
&  e-fold
&  + 818 (13.1\%)
&  0 (0\%)
&  0 (0\%)
&  0 (0\%)
&0 (0\%)
\\ \hline  
         Seed 2
&  Adaptive Learning Rate
&  + 95 (1.5\%)
&  + 766 (12.1\%)
&  + 407 (5.7\%)
&  + 28 (0.4\%)
&+ 1 (0\%)
\\ \hline  
         1
&  Factor Pruning
&  Failed to run
&  
&  
&  
&
\\ \hline  
         2
&  Dynamic Factor Adjustment
&  + 538 (8.6\%)
&  0 (0\%)
&  0 (0\%)
&  0 (0\%)
&0 (0\%)
\\ \hline  
         3
&  Green Metrics
&  + 168 (2.7\%)
&  + 97 (1.5\%)
&  + 162 (2.5\%)
&  + 176 (2.6\%)
&-18 (0.3\%)
\\ \hline  
         4
&  Quantized SGD
&  + 1173 (18.7\%)
&  -36 (0.5\%)
&  + 17 (0.2\%)
&  + 258 (3.5\%)
&+ 169 (2.2\%)
\\ \hline  
         5
&  Time Aware Early Stopping
&  + 392 (6.3\%)
&  + 4 (0.1\%)
&  0 (0\%)
&  0 (0\%)
&0 (0\%)
\\ \hline  
         6
&  Interaction Priority
&  Failed to run
&  
&  
&  
&
\\ \hline  
 7
& Micro Batching
& Failed to run
& 
& 
& 
&
\\ \hline  
 8
& Sparse Aware SGD
& + 524 (8.4\%)
& + 1 (0\%)
& 0 (0\%)
& 0 (0\%)
&0 (0\%)
\\ \hline  
 9
& Cluster Aware SGD
& Failed to run
& 
& 
& 
&
\\ \hline  
 10
& Hybrid MF
& Failed to run
& 
& 
& 
&
\\ \hline 
    \end{tabular}
    \caption{Net change in character count (sum of additions and deletions) compared to the previous iteration. The baseline implementation contained 6,260 characters across 225 lines of code.}
    \label{tab:character-changes}
\end{table*}

The AI~Scientist struggled with both implementation and methodological correctness. It failed to execute five of twelve ideas due to unresolved coding errors, often cycling through flawed versions of the same code. Even when producing executable code, outputs were frequently unreliable. For instance, in our e-fold cross-validation experiment, the AI~Scientist intended to test \( e = \{2,3,4,5\} \) but erroneously kept \( e \) fixed at 2, invalidating results. It also failed to re-run the baseline (k-fold), making e-fold cross validation appear superior -- an impossible outcome because e-fold cross-validation by design cannot lead to better performance than k-fold cross validation but only energy savings (with, ideally, comparable performance) \cite{Mahlich2024,Beel2024c,Baumgart2024}. There are many more similar examples that are available in our GitHub repository. 

These issues reveal a fundamental limitation: the AI~Scientist cannot critically assess its own results. It fails to detect methodological flaws or logical inconsistencies, making it unsuitable for autonomous scientific inquiry. Without human oversight, its findings could mislead researchers.

\subsection{Manuscripts}
The AI~Scientist generated manuscripts for all seven successfully conducted experiments, with lengths ranging from six to eight pages (median: seven). Most manuscripts (6 of 7) included tables or charts, though quality varied. References were scarce, with a median of five references per paper (range: 2–9), mostly outdated; only five (14.7\%) of the total 34 references were from 2020 or later. Foundational works, such as Goodfellow et al.'s \textit{Deep Learning} textbook, were frequently cited, suggesting weak retrieval of relevant literature.

Structural issues were common: four manuscripts (57\%)contained missing or misplaced figures, incomplete sections (e.g., ``Conclusions Here'' placeholder), duplicates, or repeated figures. Related work sections were particularly poor in all manuscripts, often with irrelevant citations. In our e-fold cross-validation experiment, the AI~Scientist failed to cite existing papers on the topic that used exactly the same terminology (e-fold cross validation), despite their availability on Semantic Scholar\footnote{\url{https://www.semanticscholar.org/search?q=\%22e-fold\%20cross\%20validation}}.

Results sections were often inadequate or misleading. Metrics were presented in plain text or simple charts, with no confidence intervals or p-values. Four of seven manuscripts (57\%) contained incorrect or hallucinated numerical results, with discrepancies in hyperparameters and performance metrics. In one case, the AI Scientist claimed to have successfully developed an energy-efficient machine learning algorithm, supposedly evidenced by better RMSE and reduction of training time from 116 to 115 seconds -- but increased memory usage. How an algorithm that is optimized for energy usage (or training time) may lead to better RMSE was not discussed (one would expect the opposite). Neither was discussed if the one-second-improvement (or 1\%), in training time was statistically (or practically) significant; nor was the implication discussed that memory usage increased. Similar logical inconsistencies appeared in other experiments.

Given these flaws, these manuscripts would likely be rejected at reputable conferences or journals. Even at lower-tier venues, the lack of proper citations and unreliable results would be problematic. However, an undergraduate submitting such work for a course project might still pass if the instructor does not scrutinize the code or experiments. Therefore, in our opinion, the AI~Scientist's output resembles that of a student rushing to complete an assignment -- formally structured but careless in execution.


\subsection{Review Functionality}
The AI~Scientist includes a reviewer agent for AI-assisted peer review, evaluated by us on both AI-generated and human-written manuscripts. To generate a review, the~AI~Scientist extracts text from a provided PDF but cannot interpret figures, tables, or supplementary material. It outputs a structured review (\texttt{review.txt}) containing a summary, identified strengths and weaknesses, improvement questions, numerical ratings for originality, clarity, significance, soundness, presentation, and contribution, plus a binary accept/reject recommendation.

\paragraph{Reviews of AI-Generated Manuscripts}
For all seven manuscripts generated by the AI~Scientist, its own review system recommended rejection. One common critique was that the research was based on a single, small dataset (MovieLens 100k). This criticism is justified, and the use of the single dataset not an intrinsic flaw of the AI~Scientist itself, but rather a consequence of our experimental pipeline, which only included this dataset. Had our pipeline contained multiple datasets, the AI~Scientist likely would have incorporated them. That being said, in an ideal scenario, the AI~Scientist would be capable of retrieving and utilizing additional datasets autonomously, rather than being entirely dependent on user-supplied input.

The AI~Scientist highlighted more valid concerns, such as weak theoretical justification and potential biases, but overlooked critical flaws, including redundant text, formatting errors, missing sections, and flawed experimental results. Most notably, the reviewer agent did not identify any of the truly serious issues that we highlighted in the sections above.

While human reviews vary in quality -- from superficial comments to detailed critiques -- the AI~Scientist’s reviews were consistently structured, though lacking depth. Despite this, they appeared plausible enough that editors and conference chairs might not immediately recognize them as AI-generated. While less rigorous than strong human reviews, they were still more substantive than the brief, low-effort reviews that sometimes occur in the conference reviewing process.

\paragraph{Reviews of Human-Written Manuscripts}
We tested\footnote{To accomplish this, we modified the AI~Scientist; this modification is available on our GitHub account.} the AI~Scientist’s reviewing capabilities with ten human-written papers on OpenReview.net\footnote{\url{https://openreview.net/}}, five that were accepted and five rejected. Reviews created by the AI~Scientist followed the same structured format as for AI-generated manuscripts, including summaries, critiques, improvement suggestions, and numerical ratings.

While indicative, the results were striking: the AI~Scientist rejected 9 out of 10 papers, including four that had been accepted by human reviewers, while recommending acceptance for only one, which had been rejected on OpenReview. This suggests a strong conservative bias and misalignment with human judgment. Despite this, its reviews raised valid concerns about methodology and biases but lacked depth, often missing broader contributions and argumentation.

Our evaluation of the AI Scientist reviewer agent, illustrates a challenge: how to evaluate AI-generated reviews, given the inconsistencies of human peer review? Should human judgment be the benchmark? Peer review is known for subjectivity, reviewer disagreements, and biases. An AI system that diverges from human reviewers may still highlight valid weaknesses or apply overly rigid criteria.

This raises another fundamental question: should AI peer review aim to mimic human reviewers or provide a more systematic alternative? If the former, the AI~Scientist falls short. If the latter, human evaluation alone may be an insufficient benchmark. A better assessment would compare AI-generated reviews against expert evaluations that measure consistency and validity.

Ultimately, while structured and plausible, the AI~Scientist’s reviews lack the depth needed for academic evaluation. It follows predefined templates but struggles with context-dependent judgment. Without improvements in assessing contributions, argumentation, and methodology, it remains better utilized as feedback tool, rather than a peer-review replacement.

\subsection{Costs and Effort}

When assessing the AI~Scientist's potential, cost is a crucial factor — and surprisingly, it remains low. The total cost\footnote{We report only costs of OpenAI's API; other costs such as electricity for our laptop or university cluster are ignored} for our main experiments, which included generating 10 ideas, running experiments for 7 ideas (from 12 total, including 2 seed ideas, of which 5 failed), and producing 7 manuscripts with reviews, amounted to only \$42 USD — an average of \$6 per manuscript. Given that our experiments were less complex than typical recommender systems research or those conducted by Sakana, their claim that a full research paper can be produced for around \$15 seems realistic.

However, human labor is still required. Setting up the AI~Scientist took approximately 5 hours, while implementing the experimental template required an additional 15 hours. It must be noted that if we were to re-implement similar pipelines, the required time would decrease due to the learning curve. If we were to implement more complex experiments, however, time would increase. Beyond this, we spent about 10 more hours brainstorming initial seed ideas, reviewing outputs, and refining results. Excluding the initial setup time, this totals 25 hours of human effort — approximately 3.5 hours per manuscript — mainly from an undergraduate student.

In summary, producing a research paper — even of relatively low quality — at an average cost of \$6 to \$15 and 3.5 hours of human effort, is remarkable. By comparison, we estimate that an undergraduate student would require at least 20 to 40 hours, while an experienced researcher — who would likely be reluctant to produce work of similar quality — might take 10 to 20 hours. Thus, the AI~Scientist operates approximately 3 to 11 times faster than a human researcher at negligible costs.

\subsection{Limitations}
Our own evaluation had limitations: we used a single dataset (MovieLens-100k) and one research domain (Green Recommender Systems), relied on fixed seed ideas and one template. While these constraints ensured a controlled assessment, they may affect generalizability and reproducibility. Despite this, we believe our findings remain meaningful.

\section{Future Work and Recommendations}

We outline several actionable recommendations for the Information Retrieval (IR) community, emphasizing practical steps for researchers, conference organizers, journal editors, and institutions. While these recommendations are primarily targeted at IR, they also reflect broader best practices in academia.

\subsection{Researchers should familiarize themselves with research-assistive AI tools}

At present, tools like the AI~Scientist hold little practical value. In contrast, more general AI tools, such as ChatGPT for text generation and GitHub Copilot for programming, already provide substantial benefits in research, improving writing, coding, and brainstorming efficiency. Researchers who ignore these readily available AI-powered assistants are already at a disadvantage compared to their peers who integrate them into their workflows.

Looking ahead, tools like the AI~Scientist will become vastly more powerful and their use may be ubiquitous. Unlike today’s prototypes, we forecast that future iterations will reliably generate research ideas, execute complex experiments, and draft high-quality manuscripts with minimal human oversight. Notably, such research-assistive AI capitalize on the underlying emergent capabilities and rapid performance progress of their underlying foundation models.

Early-career researchers, in particular, must closely monitor these developments. The use of AI in research will soon become indispensable. Those who fail to adopt AI tools risk falling behind their peers, akin to a researcher today relying on a typewriter and calculator instead of adopting a computer with text editor and programming language support. The future of research belongs to those who embrace AI-enhanced workflows and integrate them into their daily scientific practice.

The academic community must prepare for this shift and ideally lead the discussion on how AI should be integrated into scientific research. To do so, researchers should actively engage with tools like the AI~Scientist, follow advancements in this domain, and, where possible, contribute to their development and define their appropriate use. By understanding and shaping these tools, researchers ensure that AI development aligns with rigorous scientific principles rather than being dictated solely by industry interests.

\subsection{Updated Guidelines for Authors and Researchers on the Use of AI}

The ACM and other publishers provide guidelines on using generative AI in manuscript preparation. Currently, ACM permits AI to assist in manuscript writing, including generating text, images, tables, and code\footnote{\url{https://www.acm.org/publications/policies/frequently-asked-questions\#h-can-i-use-generative-ai-software-tools-to-prepare-my-manuscript}}. However, the potential for AI to conduct research and autonomously generate entire manuscripts and experimental workflows has not yet been addressed by these policies.

We urge the IR community and publishers, such as ACM, to initiate discussions on how AI-driven research tools, like the AI~Scientist, should best be integrated into scientific workflows. Specifically, the community should establish policies as to what extent AI-generated research can be submitted to conferences such as SIGIR.

At present, our assessment suggests that tools like the AI~Scientist are not yet reliable enough to contribute meaningfully to research. Consequently, manuscripts generated entirely by AI should not be accepted for publication at this time. However, this stance will need to be revisited regularly as AI capabilities evolve rapidly.

We recommend that the academic community, particularly editors of journals and members of steering or ethics committees, take an active role in revising guidelines regarding the use of AI tools in research. These revisions should be developed in collaboration with experts who possess in-depth knowledge of the capabilities and limitations of AI systems, such as the AI Scientist. Additionally, input from stakeholders like peer reviewers, program chairs of conferences, and other members of the academic publishing ecosystem will be essential to ensure that these tools are integrated effectively while maintaining academic integrity and quality control.

\subsection{Addressing the Risks of AI-Generated Assignments and Homeworks}

The ability of AI tools such as the AI~Scientist to generate well-structured academic papers that are hard to distinguish from student submissions poses an immediate challenge for universities. These systems can produce coherent, plausible reports that, especially under time constraints, may not be easily recognized as AI-generated by instructors. This presents a fundamental risk to academic integrity, as students can now generate entire assignments without engaging in the learning process. The issue is no longer speculative—AI-generated work is already at a level where instructors may fail to detect it without in-depth scrutiny. It is also difficult to prosecute, as the stochastic nature of such tools produce diverse output that is difficult to attribute back to a plagiarised source. Universities must therefore act urgently to adapt their assessment methodologies and reinforce academic standards.

A first step is to rethink assignment design, prioritizing evaluations that require learners to demonstrate personal mastery of the subject matter. This can be through iterative work, personal reflection, or in-person verification. Traditional take-home essays should be supplemented with oral examinations, in-class assignments, and project-based assessments where students demonstrate an evolving understanding of the subject, which adds value beyond what AI tools yield as a first draft. Additionally, universities should establish clear policies on AI usage, distinguishing acceptable from unacceptable applications of these tools and incorporating AI literacy into curricula. Educating students on the ethical implications and potential academic consequences of AI misuse can encourage responsible engagement. Furthermore, while AI-detection tools remain unreliable, institutions should implement verification mechanisms such as requiring students to submit intermediate drafts, video explanations, or interactive discussions about their work. These measures, when combined with updated academic integrity policies and structured instructor training, can help mitigate the risks posed by AI-generated assignments and uphold the value of university education.


\subsection{Benchmarking AI-Driven Research Agents in Information Retrieval}

We propose developing structured benchmarks to evaluate AI-driven research agents, recognizing the need for broader community discussion to establish consensus on assessment criteria. Such benchmarks are essential to determine when AI agents achieve ``Scientific AGI,'' the point at which they can autonomously conduct research at human-level quality.

A critical area is AI-powered peer review. These systems generate structured reviews, yet their reliability remains untested. Benchmarks could include expert annotations, inter-reviewer agreement analyses, and alignment with evaluation criteria to compare AI and human reviews.

AI-driven idea generation also requires assessment. Evaluations could compare AI-generated ideas with expert proposals, analyzing acceptance rates and citation impact. Experimental AI agents should be benchmarked for their ability to implement methodologies, configure parameters, and ensure reproducibility. Failure analysis is crucial to detect methodological flaws.

Manuscript generation benchmarks should go beyond fluency, assessing argument structuring, literature integration, and empirical accuracy. Automated inconsistency detection, including hallucinated citations, is essential. Blinded peer review comparisons and authorship attribution frameworks may help distinguish AI contributions.

Other AI-assisted tasks, including literature review and dataset selection, require evaluation. AI benchmarks could include retrieval relevance metrics, citation analysis, and bias detection to meet expert standards.

Developing a comprehensive benchmarking suite requires collaboration across academia, industry, and publishers. Standardized datasets, evaluation protocols, and performance metrics would help ensure scientific rigor and meaningful AI integration into research. Establishing these benchmarks will provide clear indicators of AI progress toward autonomous, human-level scientific inquiry.

\subsection{Release of AI Research Logs for Transparency}

A key advantage of AI-generated research is its ability to systematically log every step of the process. Unlike human researchers, AI systems can capture all experimental iterations, hyperparameter adjustments, and decision rationales in real time. To maximize transparency and reproducibility, AI-generated studies should include execution logs detailing raw input prompts, intermediate code modifications, and abandoned experimental runs. These logs would enable human and AI researchers to trace the AI’s reasoning process, identify potential errors, and refine methodologies.

Comprehensive versioning of all research artifacts should be mandatory. Every change to code and experimental setups should be stored in structured repositories, such as Git with detailed commit histories. Timestamped metadata should accompany these changes, documenting the AI model, system configuration, and key decision points. This will allow researchers to track the evolution of AI-generated research and compare different experimental approaches systematically.

To further improve reproducibility, AI-generated research should be accompanied by containerized environments encapsulating dependencies and configurations. Using tools like Docker or Singularity, researchers should provide prebuilt containers that replicate the AI’s computational environment. Additionally, automated reports summarizing why specific methodologies were selected or discarded would provide insight into the AI’s decision-making process, helping both human researchers and future AI agents build on existing work more effectively.

\subsection{Pilot Projects and Competitions}

The IR community can learn from initiatives such as ICLR 2025 and NeurIPS, which have explored generative AI’s role in peer review and manuscript submission through pilot projects. Expanding on these efforts, targeted pilot projects can further investigate AI’s impact on scientific research.

One promising initiative is a competition assessing AI-generated research. We propose integrating such a competition into a major conference like SIGIR, where AI-generated papers would be evaluated alongside human-authored research through a blind peer-review process. This would provide an unbiased comparison of AI's ability to conduct rigorous scientific inquiry.

While current AI research tools remain limited, a competition of this nature could become viable within the next few years. Early planning and discussion within the community would ensure that appropriate evaluation frameworks and ethical considerations are established ahead of time, enabling a structured and fair assessment of AI-driven scientific contributions.

\subsection{AI-Generated Content Tagging: Research Attribution Markup Language (RAML)}

As AI-generated research becomes more prevalent, distinguishing between human-authored and AI-assisted content is critical for transparency, academic integrity, and reproducibility. Existing solutions for image watermarking and document metadata do not adequately address the need for systematic attribution in text-based research outputs. Therefore, we propose the Research Attribution Markup Language (RAML), a standardized method to tag AI-generated content within academic manuscripts.

RAML is a JSON-based schema embedded within research papers, either as inline annotations or in a separate metadata file. It enables authors and reviewers to track which sections of a manuscript were generated, modified, or influenced by AI.

RAML defines various levels of AI involvement: Generated, meaning the content was fully created by AI; Edited, where the content was generated by AI but reviewed and revised by a human; and Suggested, in which the content was suggested by AI but primarily authored by a human. RAML also includes metadata on the AI model used, such as GPT-4, Claude, or Gemini, along with version and parameters. Additionally, it supports version control by storing AI-generated drafts, change logs, and prompts used.

To encourage adoption, RAML should be incorporated into academic publishing standards such as ACM, IEEE, and Springer, which should establish guidelines requiring RAML annotations in AI-assisted papers. Preprint servers like arXiv and OpenReview should support RAML metadata for AI disclosure. Journals and conferences should require RAML compliance for submissions involving AI. Digital Object Identifiers (DOI) should include RAML information for research reproducibility.

RAML provides a structured approach to tracking AI contributions in research manuscripts, promoting transparency and accountability. As AI plays an increasing role in scientific discovery, standardized attribution mechanisms will be essential to maintaining the credibility and integrity of academic publishing.
  
\subsection{Research Process Markup Language (RPML)}
The AI~Scientist faces difficulties in tracking detailed experimental metadata. Modifications made by its LLM-based coding assistant (Aider) are often undocumented, leading to reproducibility issues and making it hard to trace experimental changes.

We propose developing a standardized Research Process Markup Language (RPML) that captures all aspects of the research workflow in a structured format. RPML—based on a schema such as JSON or XML—should record experiment setups, code versions, container images, dataset versions, hyperparameters, and citation contexts automatically, ensuring full traceability and reproducibility.

\subsection{Standardized Dataset Annotation and Integration}
Tools like the AI~Scientist rely on user-supplied datasets that often lack detailed, standardized metadata, limiting its ability to autonomously select high-quality relevant datasets for experiments.

We propose to encourage dataset providers (e.g., Hugging Face, Kaggle, UCI Machine Learning Repository) to adopt a standardized annotation schema. This schema should include comprehensive metadata such as domain applicability, data quality scores, versioning, benchmark results, and recommended use cases. With uniform metadata, the AI~Scientist can automatically query and integrate datasets that best fit the experimental requirements. 

\subsection{Code Repositories Integration}
The AI~Scientist lacks access to recent baseline implementations from the state-of-the-art literature, making it challenging to benchmark its innovations against current best practices.

We propose to establish direct integrations with platforms such as PapersWithCode, GitHub, and Code Ocean. Standardized APIs and metadata from these platforms can be used to retrieve relevant source code repositories and performance metrics, allowing the AI~Scientist to automatically download, run, and compare these baselines against its own results.

\subsection{Participatory Strategic Retreat}
The ideas presented in our paper should be seen as an initial stimulus for discussion. To explore the future of generative AI in IR research, key stakeholders in the community must be involved—particularly leading experts in the IR field, as well as young researchers whose future will be significantly shaped by generative AI. We believe that a strategic, multi-day workshop or retreat --- such as ones held by the Dagstuhl or NII Shonan seminar series --- would be the right setting for engendering an initial, larger-scale discussion where such in-depth discussions can inform the writing of white papers that codify opinions, similar to our opinions here.  
Such papers can then help catalyse the community to comment on and organize significant IR efforts, in practice and research.

\section{Discussion and Conclusion}

Our study confirms that the AI~Scientist represents a significant step toward automating scientific research, yet our evaluation reveals that its current capabilities remain far from fulfilling its ambitious promises. While the AI~Scientist can generate research ideas, it only sometimes successfully conducts experiments, and many of its generated ideas are not truly novel. It heavily depends on user input, struggles with methodological soundness, and lacks the ability to critically assess its own results. In its present form, the AI~Scientist is, at best, an advanced research assistant that requires a lot of supervision, rather than an independent scientific agent.

A key takeaway from our study is that we believe AI systems like the AI~Scientist will meaningfully contribute to the scientific critique process in the near future, benefiting both authors and reviewers. Researchers could leverage AI-generated feedback to iteratively refine their work through adversarial loops in which an AI challenges their hypotheses, methodologies, and analyses. Similarly, peer reviewers could use AI-assisted prompts to enhance their critiques, ensuring consistency and depth. Some academic conferences, such as ICLR, have already incorporated AI-generated review suggestions into their workflows. However, our findings indicate that AI-generated reviews by the AI~Scientist often focus on surface-level critiques, while failing to detect deeper methodological flaws. This may make the job of second-level reviewers (such as area chairs or meta-reviewers) more important, as AI-generated reviews may lack depth but be well-formatted and directly address the review form's criteria. Such consolidators will have to look beyond surface-level analyses to judge whether a work holds promise.

Another promising role for the AI~Scientist lies in replication and validation. Reproducibility is a cornerstone of scientific credibility, as highlighted by recent initiatives in recommendation systems (e.g., the Best Paper Award at RecSys) and NLP (e.g., ReproHum initiatives). AI could be employed to establish proof of work, ensuring that findings are replicable and that experimental procedures are verifiable. To achieve this, AI-driven replication efforts must prioritize transparency through open-weight and open-data models. Additionally, AI can support automated metadata creation and tagging (e.g., Dublin Core, Open Data Initiative) and facilitate structured data deposits in version-controlled repositories, providing verifiable provenance for scientific claims. These mechanisms  incentivize transparent and reproducible research, potentially leading to new prestige metrics or blockchain-based certification systems for replication integrity.

Despite these potential benefits, integrating AI into scientific research presents ethical and practical challenges. AI models inherit biases from historical data and cannot independently distinguish between scientific quality and consensus. This limitation raises concerns that AI tools might reinforce outdated methodologies or amplify biases in peer review and publishing. Additionally, junior researchers may develop an overreliance on AI-generated suggestions, leading to automation bias that stifles innovation and independent thought. Research further indicates that AI can enhance scientific productivity, as shown in \cite{TonerRodgers2024}, yet its impact on researcher satisfaction remains complex. In one study, while AI-assisted workflows increased efficiency, 82\% of scientists reported lower job satisfaction afterwards. This paradox exemplifies a broader dilemma—if AI can outperform humans in key research tasks, what remains for human scientists to do? Addressing these risks requires AI-assisted workflows that actively encourage users to critically reassess assumptions and explore alternative perspectives.

One pressing concern is the potential for mass AI-generated paper submissions. Tools like the AI~Scientist could enable researchers to generate large volumes of seemingly novel but low-value papers by tweaking existing codebases and experimental parameters. Such submissions could overwhelm academic venues, burdening already-overworked reviewers. The precedent set by AI-generated nonsense papers being accepted into conferences, such as those produced by SciGen, demonstrates the urgency of this issue. Unlike SciGen, modern AI-generated papers are often indistinguishable from human writing, making them harder to detect. Without proactive measures, the academic publishing ecosystem risks being inundated with research that adds quantity but not quality. To mitigate this risk, improved authorship verification, AI disclosure policies, and systematic detection methods must be developed.

The same risks extend to students using the AI~Scientist to generate academic assignments. AI-generated essays, reports, and research papers could bypass the learning process, allowing students to submit work that appears well-structured but lacks genuine understanding or critical engagement with the subject matter. This raises significant concerns about academic integrity, as automated content generation makes plagiarism detection more difficult and blurs the line between acceptable assistance and unethical practices. Moreover, reliance on AI for coursework could hinder students' ability to develop essential skills in reasoning, analysis, and problem-solving. To address this, educators must adapt assessment strategies, incorporating oral defenses, iterative feedback loops, and AI-detection tools to ensure that students engage meaningfully with their work rather than merely outsourcing it to AI.

Looking ahead, the AI~Scientist and similar tools represent an extraordinary technological leap. AI systems that once struggled to generate coherent text can now propose research ideas, design experiments, and draft manuscripts. While the AI~Scientist currently operates at the level of an unmotivated undergraduate student, continued advancements could soon enable AI to produce research indistinguishable from that of dedicated Master’s or PhD students. Future iterations may integrate real-time literature retrieval, deeper theoretical reasoning, and more rigorous experimental validation, progressively narrowing the gap between AI-driven and human-led scientific discovery. Recent developments such as Google's\footnote{\url{https://blog.google/products/gemini/google-gemini-deep-research/}} and OpenAI's\footnote{\url{https://openai.com/index/introducing-deep-research/}} \textit{Deep Research} indicate that AI-driven research automation is rapidly advancing in this direction.

Finally, the Information Retrieval (IR) community must actively engage with these developments. The AI~Scientist relies on IR techniques for literature search, data retrieval, citation analysis, and experimental design. As AI-driven research automation grows, the IR community has the opportunity to shape its methodologies. Addressing limitations in retrieval quality, document ranking, and automated citation analysis will be critical to ensuring AI-generated research is both credible and impactful.

In summary, while the AI~Scientist falls short of its grand claims, it offers a glimpse into the future of AI-driven scientific discovery and towards ‘Artificial Research Intelligence’ (ARI). With responsible development and integration, AI could enhance knowledge generation, improve reproducibility, and streamline the research process in ways we are only beginning to explore. However, realizing this potential requires careful consideration of AI’s limitations, ethical challenges, and long-term implications for the scientific community. The time to act and to participate in evaluating, exploring, discussing and developing AI-driven research agents is now!

\section{Acknowledgements}
We used generative AI (ChatGPT) to improve this manuscript, namely for thorough proofreading and brainstorming \cite{Beel2024a}.

\bibliographystyle{ACM-Reference-Format}
\bibliography{references-ais}

\end{document}